\documentclass[aps,prb,reprint,twocolumn,longbibliography,superscriptaddress,draft=false]{revtex4-2}
\usepackage{hyperref}
\hypersetup{colorlinks=true,
		    linkcolor=blue,
		    citecolor=blue,
		    allcolors=blue}
\usepackage{natbib}
\usepackage{amsmath, amsfonts, bm, bbm, braket, mathtools}
\usepackage{array}
\usepackage{graphicx,overpic}
\graphicspath{{./}}

\usepackage[normalem]{ulem}

\usepackage{pifont}
\usepackage{lineno}

\renewcommand{\H}{\mathcal{H}}

\newcommand{\mytexttilde}{\raise.17ex\hbox{$\scriptstyle\sim$}}

\begin{document}

\title{Thermal and quantum fluctuations in the extended Kitaev-Yao-Lee
    spin-orbital model}
\author{Jiefu Cen}
\affiliation{Department of Physics, University of Toronto, Toronto, Ontario, Canada M5S 1A7}
\author{Hae-Young Kee}
\email[]{hykee@physics.utoronto.ca}
\affiliation{Department of Physics, University of Toronto, Toronto, Ontario, Canada M5S 1A7}
\affiliation{Canadian Institute for Advanced Research, CIFAR Program in Quantum Materials, Toronto, Ontario, Canada, M5G 1M1}
\begin{abstract}
Building upon the spin-1/2 Kitaev model on a honeycomb lattice, the Yao-Lee spin-orbital model provides exactly solvable quantum spin liquids with potentially better stability against perturbations due to the additional degree of freedom. 
Recently, the microscopic mechanism underlying the Yao-Lee interaction in honeycomb materials has been uncovered, leading to an extended Kitaev-Yao-Lee spin-orbital model when the celebrated Kugel-Khomskii interaction is included.  
Numerical studies of this model have identified various disordered phases, including a broad region of the nematic phase that is reminiscent of a spin-orbital liquid. Here, we investigate the origin and stability of this nematic phase via thermal and quantum fluctuations using classical Monte Carlo simulations and a generalized spin wave theory appropriate for the spin-orbital model. 
We demonstrate that the additional spin-orbital degree of freedom gives rise to strong thermal and quantum fluctuations in spin-orbital models, providing insight into the emergence of disordered phases.
\end{abstract}
\maketitle

\section{Introduction}
Frustration in magnetic Mott insulators can lead to exotic magnetic phases, such as quantum spin liquids (QSLs) \cite{ANDERSON1973153, Balents_Nature2010, Savary_RepProgPhys2016, Zhou_RMP2017, Wen_RMP2017, hermanns2018physics, Knolle_AnnRevCondMatPhys2019, Takagi_NatRevPhys2019, Broholm_Science2020}. The exemplar frustration due to the bond-dependent nearest-neighbor Ising interaction in the Kitaev model provides an exactly solvable QSL with Majorana fermions and $\bm{\mathrm{Z}}_2$ vortex excitations \cite{Kitaev_AnnPhys2006}. The Kitaev QSL has attracted much interest since the microscopic mechanism to generate the Kitaev interaction
was uncovered in materials with large spin-orbit coupling \cite{Jackeli2009PRL, Rau_PRL2014, Rau2016ARCMP, Motome2020JOP}. However, it is challenging to achieve the Kitaev QSL in solid-state materials because the fragile QSL is susceptible to other symmetry-allowed interactions that lead to conventional magnetic orders \cite{Rau_PRL2014,winter2016challenges,Liu2018PRB, Liu2020PRL, Trebst_PhysRep2022, Liu2023PRB, Rouso2024RoPP, matsuda2025rmp, Maksimov2025}.

Motivated by the Kitaev model \cite{Kitaev_AnnPhys2006}, Yao and Lee proposed a generalized model on a decorated honeycomb lattice, whose low energy effective theory can be described by both spin and pseudospin degrees of freedom on a honeycomb lattice\cite{Yao2011PRL}:
\begin{equation}
\H_{YL} = J \sum_{\langle ij\rangle_\gamma}(S^\gamma_i S^\gamma_j)\otimes(\boldsymbol{T}_i\cdot \boldsymbol{T}_j),
\end{equation}
where $\boldsymbol{S}_i$ and $\boldsymbol{T}_i$ represent the spin and pseudospin, respectively. The spins interact via a bond-dependent Ising interaction, as in the Kitaev model, but are also augmented by a Heisenberg-like pseudospin interaction. 
The model is exactly solvable with three species of free Majorana fermions coupled to a background $\bm{\mathrm{Z}}_2$ gauge field. It exhibits spin-1 fermionic excitations and non-abelian vortex excitations in the presence of an external field. This model is more robust to perturbations than the Kitaev model due to the additional degree of freedom \cite{Chulliparambil_PRB2020, Seifert_PRL2020, Chulliparambil_PRB2021, Akram_PRB2023, Nica_npjQM2023, Poliakov_PRB2024, Wu_PRL2024}.

Recently, a microscopic theory to achieve the Yao-Lee (YL) type interaction has been uncovered in $d^9$ or $d^7$ ions on a honeycomb lattice, where $e_g$ orbitals serve as an additional degree of freedom \cite{Churchill_npjQM2025}.
The bond-dependent Kitaev interaction arises from the bond-dependent effective hopping mediated by the p-orbitals of intermediate anions with strong spin-orbit coupling. The Heisenberg-type interaction for the $e_g$ orbital pseudospin can be achieved through a sublattice transformation of the original XXZ-like interaction. 
When other interactions, such as the Kugel-Khomskii\cite{Kugel1973}, are included, the numerical study using the classical Monte Carlo simulation (CMC) and the exact diagonalization calculation (ED) on a small cluster (12-site) has reported that the model displays a broad region of a disordered phase named the nematic phase (NP), reminiscent of a spin-orbital liquid. 
In the classical model, this NP is characterized by the product of spin and orbital orders, exhibiting a sharp peak at the $M$-point in the spin-orbital correlation, whereas the spin and orbital orders alone are absent. On the other hand, in the quantum model, the bond energy anisotropy in the absence of a long-range spin or orbital order is observed in the NP, in contrast to the paramagnetic state, where the bond energy is uniform. 

Although numerical studies have established the existence of this phase, its emergence even in the classical limit remains unclear, since spin models in this limit typically develop magnetic orders at zero temperature. 
Here, we investigate the origin and stability of this NP via thermal and quantum fluctuations using CMC and a generalized spin wave theory (SWT) appropriate for the spin-orbital model. We find that the NP exhibits macroscopic degeneracies arising from the spin-orbit degree of freedom and size-independent low energy costs, so it emerges at low temperatures due to thermal effects. In addition, the NP is also stabilized by the significant quantum fluctuations in the generalized SWT. We consider all cubic and quartic terms up to order $\Lambda^{-1}$ (the generalized spin length $\Lambda$) in the spin wave expansion, which are important for general spin-orbital models since the cubic terms are present even at order $\Lambda^{0}$.

The paper is organized as follows. In Sec. \ref{sec_2}, we review the spin-orbital model and the characteristics of various phases. In Sec. \ref{sec_3}, we present and analyze the classical phase diagrams at zero and finite temperatures obtained by CMC simulations. We explain why thermal fluctuations favor the NP over other disordered phases at low temperatures.
In Sec. \ref{sec_4}, we study quantum effects in the phase diagram using a generalized SWT. 
In the last section, we summarize our main results and discuss open questions for future studies.

\section{Spin-orbital model and nematic phase} \label{sec_2}
 The spin-orbital model for magnetic ions that possess both $e_g$ orbital (one electron or hole) and spin degrees of freedom, surrounded by heavy anions forming an octahedral cage, was derived in Ref. \cite{Churchill_npjQM2025}.
It has been shown that the YL type bond-dependent interaction comes from the effective bond-dependent interorbital hopping between $e_g$ ions obtained by integrating out the ligand p-orbitals \cite{Stavropoulos_PRL2019}. Since there is no spin-orbit coupling at the magnetic ions, the two $e_g$ orbitals provide the additional degree of freedom. 
When direct intraorbital hopping is considered, the Kugel-Khomskii interaction \cite{Kugel1973,Kugel_SovPhys1982, Brink_PRB2001, Mostovoy_PRB2002, Khomskii_JETP2016} is also included, leading to the following spin-orbital model.
\begin{equation} \label{eq1}
\H = J \sum_{\langle ij\rangle_\gamma}( - a\boldsymbol{S}_i\cdot \boldsymbol{S}_j + 2S_i^\gamma S_j^\gamma +b)
\otimes(\boldsymbol{T}_i\cdot \boldsymbol{T}_j-b),
\end{equation}
where 
$a$ and $b$ account for the relative strength of the Kugel-Khomskii interaction and the effective interaction due to the hopping mediated through anions. The orbital pseudospin $T^x$ and $T^z$ carry quadrupolar moments, while $T^y$ carries an octupolar moment, and a sublattice transformation $T^y_i \rightarrow (-1)^iT^y_i$ has been used to transform the XXZ-like orbital interaction to the Heisenberg interaction. 
Note Ref.\cite{Churchill_npjQM2025} uses a different sublattice transformation $T^{x/z}_i \rightarrow(-1)^iT^{x/z}_i$. The resulting ferromagnetic orbital order denoted by ${\tilde{FO}}$ is equivalent to the antiferromagnetic orbital (AFO) order in the current study.

\begin{figure}
    \centering
    \includegraphics[width=1.0\linewidth]{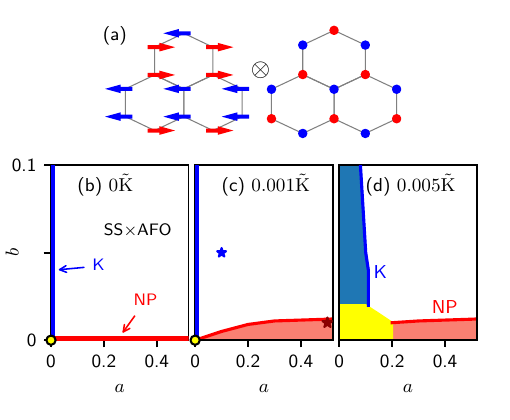}
    \caption{Classical phase diagrams of the spin-orbital model at various temperatures. The unit of temperature is $\mathrm{\tilde{K}}=k_BT/J$. (a) The stripy spin (SS) and the antiferromagnetic orbital (AFO) configurations of the ${\rm SS \times AFO}$ phase. (b) At zero temperature, when $a>0$ and $b>0$, the model is fully ordered (${\rm SS \times AFO}$) with the stripy spin and the antiferromagnetic orbital (white region). The YL point is $a=0$ and $b=0$ (yellow circle). The Kitaev phase (K) is the blue line $a=0$ and $b>0$. The nematic phase (NP) is the red line $a>0$ and $b=0$.   (c) At a low temperature $T=0.001 \,\mathrm{\tilde{K}}$, the Kitaev phase and the YL point remain unchanged, but the NP extends to $b>0$ (red region). The red and blue stars highlight the parameters used to study the finite-temperature phase transitions in Fig. \ref{Fig3}. (d) At a higher intermediate temperature $T=0.005 \,\mathrm{\tilde{K}}$, the Kitaev phase, the YL point, and the NP are extended by thermal fluctuations. 
    }
    \label{Fig1}
\end{figure}

The numerical study in Ref. \cite{Churchill_npjQM2025} employing the CMC and ED has revealed various exotic phases. The phase diagram for small $a$ and $b$ is shown in Fig. \ref{Fig1}.
When $a=b=0$, the YL spin liquid denoted by the yellow circle in Fig. \ref{Fig1}(b) is found. When $a=0$ and $b>0$ (i.e., y-axis), we get the classical Kitaev spin liquid (blue line/region in Fig. \ref{Fig1}) with the antiferromagnetic orbital (AFO) order. 
For $a>2$ and a large $b$ (not shown in Fig. \ref{Fig1}), the ordered phase with the antiferromagnetic spin (AFM) and the AFO configuration occurs, denoted as AFM$\times$AFO.
For $0<a<2$ and a large $b$, the ordered phase has the stripy spin (SS) and the AFO configuration, denoted as SS$\times$AFO (white region in Fig. \ref{Fig1}). The spin and orbital configurations are shown in Fig. \ref{Fig1}(a). 
For $0<a<2$ and a small $b$, the NP phase emerges (red line/region in Fig. \ref{Fig1}), characterized by a sharp feature at the momentum $M$-point ($\boldsymbol{q=M}$) in the spin-orbital structure factor, $ST(\boldsymbol{q})$, but without a sharp feature in the spin and orbital structure factors $S(\boldsymbol{q})$ and $T(\boldsymbol{q})$. These structure factors are defined as follows:
\begin{align} \label{eq_sf}
    ST(\boldsymbol{q}) &= \frac{1}{N^2}\sum_{ij} \langle(\boldsymbol{S}_i \cdot\boldsymbol{S}_j)(\boldsymbol{T}_i \cdot\boldsymbol{T}_j)\rangle e^{-i\boldsymbol{q}(r_i-r_j)}, \nonumber
    \\
    S(\boldsymbol{q}) &= \frac{1}{N^2}\sum_{ij} \langle(\boldsymbol{S}_i \cdot\boldsymbol{S}_j)\rangle e^{-i\boldsymbol{q}(r_i-r_j)}, \nonumber
    \\
    T(\boldsymbol{q}) &= \frac{1}{N^2}\sum_{ij} \langle(\boldsymbol{T}_i \cdot\boldsymbol{T}_j)\rangle e^{-i\boldsymbol{q}(r_i-r_j)}.
\end{align}
It was shown that the NP region is further extended to higher $b$ values for the quantum model in ED calculations.\cite{Churchill_npjQM2025}

In this study, we address two questions: how the NP arises and why it becomes significantly broader in the quantum model compared to classical simulations. Our analysis focuses on the roles of temperature and quantum effects.
To understand the emergence of the NP, we simulate and analyze the various classical ground states at zero and finite temperatures. 
Here, we will focus only on the parameter range $0<a<2$, since we are interested in the interplay between the classical Kitaev spin phase, the YL phase, and the NP, which exist in the small $a$ and $b$ regime as shown in Fig. \ref{Fig1}. For $a \geq 2$, two fully ordered phases denoted by AFM$\times$AFO and SS$\times$AFO are found. In Appendix \ref{apdx_A}, we show that the transition between them occurs at $a=2$, regardless of the finite value of $b$ in the classical model. 

\section{Classical Monte Carlo simulation} \label{sec_3}

The classical phase diagrams at various temperatures are shown in Fig. \ref{Fig1}(b)-(d). The unit of temperature is $\mathrm{\tilde{K}}=k_BT/J$, and the number of unit cells is $16\times16$. { We first analyze the zero-temperature phase diagram in Fig. \ref{Fig1}(b), then study the finite-temperature phase diagrams in Fig. \ref{Fig1}(c)-(d), obtained by CMC with parallel tempering to improve accuracy (see Appendix \ref{apdx_B}) \cite{Hukushima1996, Katzgraber2006, HAMZE2010}.}

\subsection{Zero-temperature phase diagram}

 For the SS$\times$AFO phase when $a>0$ and $b>0$, two of the three nearest neighbor bonds (denoted by $\delta$) have antiparallel spins, while the remaining bond has parallel spins, so $\sum_{\delta}\langle {\bf S}_i \cdot {\bf S}_{i+\delta} \rangle = -S^2$, i.e, the average per bond is $-S^2/3$ with the spin length $|{\bf S}_i|=S$. 
The AFO ordering leads to the uniform $\langle {\bf T}_i \cdot {\bf T}_j \rangle = -S^2$ per bond with the orbital pseudospin length $|{\bf T}_i|=S$. 
Thus, the energy of the SS$\times$AFO ($\epsilon_{\rm SS\times AFO}$) per bond in units of $JS^2$ is given by  
\begin{equation} \label{eq_E_SS}
   \epsilon_{\rm SS \times AFO} = - S^2 \left(\frac{a+2}{3} +\frac{b}{S^2} \right) \times \left( 1+ \frac{b}{S^2}\right).
\end{equation}
From now on, we omit the $JS^2$ unit.
The total energy is $\epsilon_{\rm SS \times AFO} \times 3N$, since there are $3N$ bonds where $N$ is the total number of unit cells.

\begin{figure}
    \centering
    \includegraphics[width=0.5\linewidth]{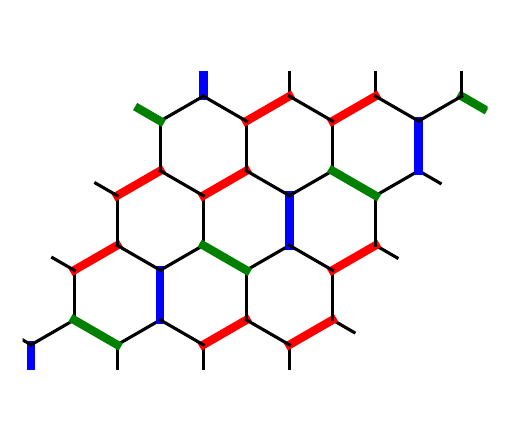}
    \caption{Example of a Cartesian state \cite{Baskaran_PRB2008} of the classical Kitaev model on a $4\times4$ lattice. The two neighbouring spins are aligned along the $x$, $y$, and $z$ axes on the red, green, and blue bonds, respectively, meaning the minimum bond energy is satisfied on these bonds. For the remaining black bonds, the energy is zero. Each unit cell has one satisfied bond.}
    \label{Fig2}
\end{figure}

For the Kitaev phase with the AFO order when $a=0$ and $b>0$, the energy $\epsilon_K$ per bond is the product of the classical Kitaev spin liquid \cite{Baskaran_PRB2008} times the AFO order, given by 
\begin{equation}
    \epsilon_{\mathrm{K}} = -S^2 \left(\frac{2}{3} +\frac{b}{S^2} \right) \times \left( 1+\frac{b}{S^2} \right).
\end{equation}
 { Since the orbital sector has the trivial AFO order, only the spin ground state manifold is considered.}
 Typical ground state configurations of the classical Kitaev spin model are the so-called Cartesian states with $\langle {\bf S}_i \cdot {\bf S}_j \rangle = S^2/3$ per bond since only one minimum bond energy is satisfied in a unit cell, while the other two are zero, as shown in Fig. \ref{Fig2}.
There are $2.76^{N}$ ways to construct Cartesian states.
In addition, infinitely many more ground states exist that interpolate any two Cartesian states \cite{Baskaran_PRB2008, Chandra_PRE2010, Rouso2024RoPP}.

When $b=0$, we observe the YL phase when $a=0$ and the NP when $a>0$, with the energies per bond given by 
\begin{equation} \label{eq_E_NP}
    \epsilon_{\mathrm{YL}}=-\frac{2}{3}S^2, \;\; \epsilon_{\mathrm{NP}}=-\frac{a+2}{3}S^2.
\end{equation}
 Clearly, $\epsilon_{\mathrm{YL}}=\epsilon_{\mathrm{K}}=\epsilon_\mathrm{NP}$ when $a=0$ and $b=0$, which means that they all have the same energy and belong to the YL ground state manifold. 
 It is important to note that the zero-temperature phase diagram is different from that in Ref. \cite{Churchill_npjQM2025}.
We believe the temperature used in the previous study was not sufficiently low, so finite-temperature effects were included. Next, we study the degeneracies of the YL phase and the NP.

When $b=0$, our model has a much larger degeneracy since only the spin-orbital quartic terms $\sum_{\langle ij \rangle}(-a{\bf S}_i \cdot {\bf S}_j + 2 S_i^\gamma S_j^\gamma)( {\bf T}_i \cdot {\bf T}_j)$ 
are present. Thus, the energy remains the same if the spin and the orbital moments at one site are flipped together. In other words, each site has the $(\boldsymbol{S}_i,\boldsymbol{T}_i)\rightarrow (-\boldsymbol{S}_i,-\boldsymbol{T}_i)$ degree of freedom, leading to { additional $2^{2N}$ macroscopic degeneracies on top of the spin-only model}. Therefore, we speculate that the ground state manifold of the YL state is that of the Kitaev spin model times $2^{2N}$, which future studies need to verify.

Similarly, we find that the NP is obtained by applying $(\boldsymbol{S}_i,\boldsymbol{T}_i)\rightarrow (-\boldsymbol{S}_i,-\boldsymbol{T}_i)$ flips to the ${\rm SS \times AFO}$ order, so
the ground state degeneracy of NP is $\sim2^{2N}$. The NP inherits the $M$-point ordering wavevector in $ST (\boldsymbol{q}=\boldsymbol{M})$ from the ${\rm SS \times AFO}$ order, since flipping $(\boldsymbol{S}_i,\boldsymbol{T}_i)$ does not change the ordering in the spin-orbital $S \otimes T$ sector, but destroys the ordering in the spin and orbital degrees of freedom, i.e. $\langle S_i^\alpha S_j^\beta \rangle = 0$ and $\langle T_i^\alpha T_j^\beta \rangle = 0$ for any $\alpha$ and $\beta$. 

\subsection{Finite-temperature phase diagram}

At finite temperature, the macroscopically degenerate excited states have larger entropies and overcome the energy costs, so the Kitaev phase, the YL point, and NP can all extend into the fully ordered $a>0$ and $b>0$ region, as shown by the phase diagram Fig. \ref{Fig1}(d). However, NP has distinct thermal effects. At a very low temperature $T=0.001\,\mathrm{\tilde{K}}$, the Kitaev phase and the YL point remain unchanged, but the NP extends to $b>0$, as shown by Fig. \ref{Fig1}(c) (note the similarity to the classical phase diagram in Ref. \cite{Churchill_npjQM2025}, confirming that the result was obtained at a tiny but finite temperature). As the temperature increases, the Kitaev phase and the YL point start to extend, but the NP does not extend further. There is no qualitative change for parameter ranges up to $0<a<2$ and $0<b<0.25$, so we zoom in around small $a$ and $b$ values. 
To understand how these phases evolve in phase space as the temperature increases, we consider how the degenerate ground states gain energy and split when $a>0$ and $b>0$. 

The Kitaev phase's energy increases since $a>0$ reduces the energy of the satisfied bonds of the Cartesian states:
\begin{equation} \label{eq_E_K}
    \epsilon_{\mathrm{K}} = -S^2 \left(\frac{2-a}{3} +\frac{b}{S^2} \right) \times \left( 1+\frac{b}{S^2} \right).
\end{equation}
Similarly, when $b=0$, we also expect the YL phase to gain energy by  
\begin{equation} \label{eq_E_YL}
    \epsilon_{\mathrm{YL}}=-\frac{2-a}{3}S^2.
\end{equation}
The contribution of $a$ is opposite to the ${\rm SS \times AFO}$ phase since the stripy phase has two satisfied bonds and one unsatisfied bond in a unit cell. 
The degeneracies of the Kitaev and the YL phase due to $a>0$ remain unchanged, since all Cartesian states gain the same energies.

When $b>0$, the situation is different for the NP. Starting from the ${\rm SS \times AFO}$ phase, each flip of $(\boldsymbol{S}_i,\boldsymbol{T}_i)$ now costs energy $2\Delta\epsilon_{st}$ per bond given by
\begin{align} \label{eq_Est}
    \Delta\epsilon_{st} = \frac{1}{N}\large[\frac{(2+a)}{3}b +b \large].
\end{align}
Note that we get the $1/N$ factor since each flip is local.
The energy cost comes from the quadratic $S_iS_j$ and $T_iT_j$ terms in the Hamiltonian, which change sign when $(\boldsymbol{S}_i,\boldsymbol{T}_i)\rightarrow (-\boldsymbol{S}_i,-\boldsymbol{T}_i)$. The degeneracy of a single flip is $2\binom{N}{1}$, as we can pick any site to flip. We can easily extend this to flipping multiple sites if we assume that the flipped sites are not near each other. For example, this can be achieved by flipping sites only in the same sublattice. In this case, the state with the energy cost of $2m \Delta\epsilon_{st}$ has the degeneracy of the order of $\binom{N}{m}$, where $m$ is the number of flipped sites. If there are nearest-neighbor flipped sites, the domains of the flipped sites must be considered, thereby reducing the energy cost and the degeneracy. 

The NP consists of states with macroscopic degeneracies and size-independent low energy costs. In contrast, although the Kitaev phase and the YL phase have much larger degeneracies, the spin configurations of the Cartesian states are very different from the ${\rm SS \times AFO}$ ground state, so the energy costs also scale with the lattice size. This key difference explains why the NP (red region) extends at low temperatures, whereas the Kitaev phase (blue region) and the YL point (yellow region) remain essentially unchanged at $T=0.001\,\mathrm{\tilde{K}}$ but expand noticeably at $T=0.005\,\mathrm{\tilde{K}}$, as illustrated in Fig. \ref{Fig1}(c)-(d).

\subsection{Static structure factor}

\begin{figure}
    \centering
    \includegraphics[width=1.0\linewidth]{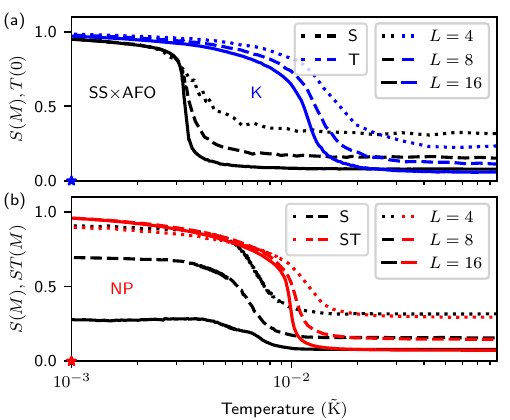}
    \caption{Structure factors of spin $S(\boldsymbol{q=M})$, orbital $T(\boldsymbol{q=0})$, and spin-orbital $ST(\boldsymbol{q=M})$ vs. temperature. The SS$\times$AFO ground state has both maximum $S(\boldsymbol{M})$, $T(\boldsymbol{0})$, and $ST(\boldsymbol{M})$. (a) When $a=0.1$ and $b=0.05$, denoted by the blue star in Fig. \ref{Fig1}(c), the Kitaev phase emerges at intermediate temperatures, characterized by the maximum $T(\boldsymbol{0})$ only. (b) When $a=0.5$ and $b=0.01$, denoted by the red star in Fig. \ref{Fig1}(c), the NP emerges at low temperatures, characterized by the maximum $ST(\boldsymbol{M})$ and a small $S(\boldsymbol{M})$. The transition from the NP to the ${\rm SS \times AFO}$ is broad and sensitive to finite-size effects due to the energy cost of the NP. See the main text for discussion. }
    \label{Fig3}
\end{figure}

The static structure factors also reflect the distinct thermal effects of the NP.
The various phases are identified by the spin, orbital, and spin-orbital static structure factors in Eq. \ref{eq_sf}. The ${\rm SS \times AFO}$ has both maximum $S(\boldsymbol{q=M})$, $T(\boldsymbol{q}=0)$, and $ST(\boldsymbol{q=M})$ (the stripy spin phase has an ordering wavevector at ${\bf q}={\bf M}$). The Kitaev phase has a maximum $T(\boldsymbol{q}=0)$. Figure \ref{Fig3}(a) plots $S(\boldsymbol{q=M})$ and $T(\boldsymbol{q}=0)$ versus temperature at $a=0.1$, $b=0.05$ (blue star in Fig. \ref{Fig1}(c)), and various lattice sizes $L$. The number of unit cells is $L\times L$. The model transitions from the ${\rm SS \times AFO}$ ground state to the Kitaev phase at intermediate temperatures before reaching the high-temperature disordered phase. 

The NP has a maximum $ST(\boldsymbol{q=M})$ but smaller $S(\boldsymbol{q=M})$ and $T(\boldsymbol{q}=0)$ that depend on the lattice size, as shown by Fig. \ref{Fig3}(b) with $a=0.5$ and $b=0.01$ (red star in Fig. \ref{Fig1}(c)).
The decrease in $S(\boldsymbol{q=M})$ versus the lattice size is consistent with the energy analysis above: the energy cost of NP is size-independent, but the entropy gain scales with the lattice size, so the larger the size, the easier it is to flip the moments at multiple sites, and the smaller $S(\boldsymbol{q=M})$.
Thus, as $L\rightarrow\infty$, we expect $S(\boldsymbol{q=M})\rightarrow 0$, $T(\boldsymbol{q}=0)\rightarrow 0$, and the energy of NP approaches Eq. \ref{eq_E_NP} above.
In addition, the temperature required to flip the moments is much lower for the NP than for the Kitaev phase.
With this model parameter, the NP is a clear intermediate phase at $L=4$. However, at a larger $L$, NP is thermally selected down to the lowest temperature $T=0.001 \,\mathrm{\tilde{K}}$ of our parallel tempering simulations.

\section{Quantum effects: generalized spin wave theory} \label{sec_4}

Spin wave theory is a standard method for including quantum effects on classical ordered phases. One can apply conventional $SU(2)$ SWT \cite{HolsteinPrimakoff} separately to the spin $S$ and orbital $T$ degrees of freedom. However, our spin-orbital model contains contributions of order $S^4$ from quartic terms such as $({\bf S}_i\cdot{\bf S}_j)( {\bf T}_i\cdot {\bf T}_j)$, as well as contributions of order $S^2$ from quadratic terms such as $({\bf S}_i \cdot {\bf S}_j)$ and $({\bf T}_i \cdot {\bf T}_j)$. To account for quantum fluctuation effects from these different orders on an equal footing, one must include higher-order corrections from the $1/S$ expansion, which goes beyond linear SWT. Furthermore, $SU(2)$ SWT does not incorporate higher multipolar fluctuations, such as spin-orbital fluctuations, as fundamental excitations. These limitations of $SU(2)$ SWT can be addressed using a generalized $SU(4)$ SWT, which treats all 15 types of interactions ($S_i^\alpha$, $T_i^\alpha$, and $S_i^\alpha T_i^\beta$) on an equal footing through a mapping onto the 15 $SU(4)$ generators \cite{Li_SU4_1998PRL, Generalized_SWT}. In this framework, the influence of quartic terms is already partially incorporated at the linear SWT level.

Here, we introduce a generalized $SU(4)$ SWT appropriate for spin-orbital models following Ref. \cite{Joshi_SU4_1999PRB}. The four basis states for the spin and orbital operators are as follows.
\begin{align}
    \ket{1} &= \ket{\frac{1}{2},\frac{1}{2}}, &\ket{2}& = \ket{-\frac{1}{2},\frac{1}{2}}, \nonumber
    \\
    \ket{3} &= \ket{\frac{1}{2},-\frac{1}{2}},&\ket{4}& = \ket{-\frac{1}{2},-\frac{1}{2}}. 
\end{align}
The spin and orbital operators are expressed as linear combinations of the $SU(4)$ generators $O_m^n$, where $O_m^n \ket{l}=\delta_{n,l}\ket{m}$ transitions between basis states and satisfies $\sum_mO^m_m=1$ and $(O^m_n)^\dagger=O^n_m$. The Lie algebra is 
\begin{equation}
[O^n_m,O^l_k] = \delta_{n,k}O^l_m - \delta_{m,l}O^n_k.
\end{equation}
Here are some examples of the mapping: 
\begin{align}
    &\sigma^z = \frac{1}{2}(O^1_1 - O^2_2 + O^3_3 - O^4_4), \nonumber\\
    &\sigma^+ = O^2_1 + O^4_3, \nonumber\\
    &\tau^+ = O^3_1 + O^4_2, \nonumber\\
    &\sigma^+\tau^+ = O^4_1,
\end{align}
where $\sigma$ and $\tau$ are the $SU(2)$ spin and orbital operators, respectively. Note the $\sigma\tau$ product terms (coming from the quartic terms in the Hamiltonian) are mapped to linear $SU(4)$ generators.
Next, the generalized Holstein-Primakoff transformation (HPT) \cite{onufrieva1985, Joshi_SU4_1999PRB} is used to map the $SU(4)$ generators to three boson operators as follows:
\begin{align} 
\label{eq_HPT}
    O^1_1 &= \Lambda - \sum_{n=2}^{4}b_n^\dagger b_n, \nonumber\\
    O^1_n &= \sqrt{\Lambda}b_n^\dagger \sqrt{1 - \frac{1}{\Lambda}\sum_{l=2}^{4}b_l^\dagger b_l}, \;\;n\ne1 \nonumber\\
    O^l_n &= b_n^\dagger b_l, \;\;\;\;\; l,n=2,3,4,
\end{align}
where $\ket{1}$ is chosen to be the local ground state at a site, and $\Lambda=1$ is the generalized spin length for the spin-orbital models. { As in $SU(2)$ SWT, the square root in Eq. \ref{eq_HPT} can be expanded in $1/\Lambda$ to include higher order magnon-magnon interactions.} Based on the transitions, we call $O^1_2\sim b^\dagger_2$ the spin excitation, $O^1_3\sim b^\dagger_3$ the orbital excitation, and $O^1_4\sim b^\dagger_4$ the spin-orbital excitation.
Using this theory, we calculate the magnon excitations of the fully ordered ${\rm SS \times AFO}$ state and study its stability in the presence of quantum fluctuations by examining the gap-closing condition of the magnons. 

Our method is applicable to general spin-orbital models with stable ordered phases.
However, this comes with a caveat inherent to linear SWT, as revealed by the omitted cubic terms. Unlike the conventional $SU(2)$ SWT, the generalized SWT on our model generates cubic terms even if we just expand Eq. \ref{eq_HPT} in the linear $\Lambda^0$ order. These cubic terms are ignored in the non-interacting Hamiltonian. To incorporate them consistently, one must proceed to the next order $\Lambda ^{-1}$ and include all contributions in that order. Below, we present the linear SWT results and the importance of higher order corrections.


\subsection{Limitation of linear spin wave theory}

By truncating the generalized HPT in Eq. \ref{eq_HPT} to the order $\Lambda^0$, a Hamiltonian quadratic in the boson operators can be obtained. We first transform the local spin and orbital operators in the unit cell of the ${\rm SS \times AFO}$ order (4 sites per magnetic unit cell) so that $\ket{1}$ is the local ground state. Then, we diagonalize the resulting quadratic $24\times24$ Hamiltonians in momentum space using standard methods \cite{COLPA1978327}.

The phase boundary of the ${\rm SS \times AFO}$ state is obtained by the gap-closing condition of the magnon spectra. We can gain insights into the phases by inspecting the magnon wavefunctions. We observe a Goldstone mode for the orbital excitation since our model (Eq. \ref{eq1}) always has a $SU(2)$ symmetry in the orbital interaction. When $a=1$, an additional Goldstone mode appears for the spin excitation since the spin part of the models can be mapped to the Heisenberg ones with sublattice transformation \cite{Chaloupka2015hidden}.

\begin{figure}
    \centering
    \includegraphics[width=1.0\linewidth]{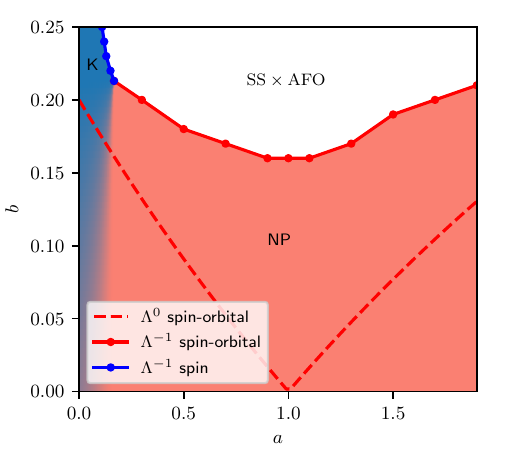}
    \caption{Quantum phase diagram of the spin-orbital model by a generalized spin wave theory at various expansion orders (the generalized spin length $\Lambda$). The theory computes the phase boundary using the magnon gap-closing condition of the fully ordered ${\rm SS \times AFO}$ phase. The Kitaev phase boundary (blue line for order $\Lambda^{-1}$) has the vanishing spin magnon, while the NP boundary (red solid line for order $\Lambda^{-1}$ and dashed line for order $\Lambda^0$) has the vanishing spin-orbital magnon.}
    \label{Fig4}
\end{figure}

Importantly, we find that the four (four sites in the unit cell) degenerate gap-closing magnons are spin-orbital excitations. 
{
Within linear SWT, the spin-orbital magnon does not interact with the spin or the orbital magnon. Also, the two sites connected by the x/y bond interact as a pair, and the other two sites do so in the same way. Therefore, it is enough to show the Hamiltonian within the spin-orbital subspace for just one pair in the basis of $\Psi^\dagger(k) =[b^\dagger_4(k),\, \tilde{b}^\dagger_4(k),\, b_4(-k),\, \tilde{b}_4(-k)]$:
\begin{equation}
    \mathcal{H}_{st}(k)=\frac{J}{4} 
    \begin{pmatrix}
    A & 0 & 0 & B(k) \\
    0 & A & B(-k) & 0 \\
    0 & B^*(-k) & A & 0 \\
    B^*(k) & 0 & 0 & A
    \end{pmatrix},
\end{equation}
where $b^\dagger_4$ and $\tilde{b}^\dagger_4$ represent the spin-orbital bosons for the two sites in the magnetic unit cell, respectively. $A=2(2+a)b+6b$ and $B(k)=(1-a)(1+e^{-ik})$. 
The diagonal terms correspond to the classical spin-orbital excitation energy $2\Delta\epsilon_{st}$ of the NP, consistent with our classical analysis above. The off-diagonal terms are quantum effects (maximum at $k=0$), leading to the phase boundary at linear order given by 
\begin{equation}
    b_c=\frac{|1-a|}{2+a+3},
\end{equation}
where $b_c$ is the critical gap-closing line shown in Fig. \ref{Fig4}.
}
Thus, the phase boundary separates the NP and the ${\rm SS \times AFO}$ phase. 
The off-diagonal pairing terms are quantum effects, so inside the NP, quantum fluctuation in the spin-orbital degree of freedom destroys the classical ${\rm SS \times AFO}$ order. 
{
Note that SWT fails for the NP, which has a higher classical energy than the ${\rm SS \times AFO}$ order, so in this sense, the NP is not a conventional ordered state stabilized by quantum fluctuations, distinct from the usual quantum order by disorder mechanism.
}

The decrease of the transition to $b=0$ when $a=1$ seems to suggest that quantum effects disappear when $a=1$. 
However, this reflects the limitation of linear SWT, evident from the neglected cubic terms. Next, we show that magnon interactions in the $\Lambda^{-1}$ order, in particular cubic terms, increase $b_c$ significantly.

\subsection{Importance of the higher order correction}
The order $\Lambda^{-1}$ expansion of Eq. \ref{eq_HPT} generates both cubic and quartic magnon interaction terms. The quartic terms are treated with the mean-field Hartree-Fock method, while the cubic terms are treated with second order perturbation to the non-interacting states \cite{Rastelli_1985, Chubukov_1994, Chernyshev2009PRB}. We are interested in the corrections to the magnon energies. 

We find that the quartic interaction terms do not change the wavefunction of the gap-closing spin-orbital magnon; i.e., they simply renormalize the non-interacting magnon energies to $\tilde{E}_{st}$ (see Appendix \ref{apdx_C}), so the transition is still characterized by spin-orbital excitations. 
The cubic terms provide a correction $\Delta E_{st}\sim -1/E_{st}$ in second order perturbation, which diverges as $b$ goes to zero, so the transition defined by $\tilde{E}_{st}+\Delta E_{st}=0$ occurs at a larger $b$ value. In other words, because magnon-magnon interactions greatly enhance the quantum fluctuations that destabilize the classical $\mathrm{SS\times AFO}$ state, the transition to the NP occurs only at much larger $b$ values in the quantum phase diagram, as indicated by the red line in Fig. \ref{Fig4}. 

Furthermore, we find another transition from the classical ${\rm SS \times AFO}$ order to a disorder state, denoted by the blue line in Fig. \ref{Fig4}. This transition is characterized by a gap-closing magnon in the spin degree of freedom, consistent with the Kitaev spin phase with an AFO order.
The transition between the Kitaev phase and NP remains unclear because the SWT applies only to the ${\rm SS \times AFO}$ phase, so only the phase boundary is determined.

In this generalized SWT, the cubic terms indeed significantly enhance quantum fluctuations and produce a phase boundary at large $b$ values, consistent with the ED results in Ref. \cite{Churchill_npjQM2025}. { Previous studies with $SU(2)$ and $SU(3)$ SWT have reported similar effects of magnon-magnon interactions: state‑independent renormalization from quartic terms and substantial energy corrections from cubic terms} \cite{Chernyshev2014_PRL, Mendili2025}.

\section{Summary and Discussion}
In summary, we study how the spin-orbital nematic phase emerges in the extended Kitaev-Yao-Lee model via thermal and quantum fluctuations. The NP is characterized by a finite spin-orbital structure factor, originating from excitations in the spin-orbital degree of freedom.
The classical phase diagrams at various temperatures show that the NP is stabilized despite not being the ground state of the model, because it consists of the macroscopic degenerate low-lying excitations with size-independent costs from the ${\rm SS \times AFO}$ ground state.
When quantum fluctuations are included using a generalized SWT for spin-orbital models, the NP is significantly enhanced. Our SWT includes quartic and cubic terms up to order $\Lambda^{-1}$. The effects of higher-order interaction terms on a larger system size need further investigation, especially since the NP can host many spin-orbital excitations. 


Our generalized SWT results reveal strong quantum effects in the disordered phases of both the Kitaev and the NP regions. 
The use of alternative theoretical frameworks, such as the $SU(4)$ Schwinger boson approach \cite{Arovas1988PRB, auerbach2012interacting, Zhang2022PRB}, to investigate nematic and other disordered phases in spin‑orbital systems is an excellent project for future study.

{ Our study suggests that materials with strong spin-orbital interactions, such as 3d$^7$ or 3d$^9$ compounds hosting one electron or hole in a two-orbital manifold and exhibiting weak or quenched spin–orbit coupling at the transition-metal sites, are promising candidates for realizing stable spin–orbital liquids.}

\section{Acknowledgments} 
This work is supported by the NSERC Discovery Grant No. 2022-04601 and NSERC CREATE program No. 575280-2023. H. Y. K. acknowledges support from the Canada Research Chairs Program No. CRC-2019-00147.
This research was enabled in part by support provided by Compute Ontario,  Calcul Québec, and the Digital Research Alliance of Canada.

\appendix

\section{transition between fully ordered phases} \label{apdx_A}

The ground state energy of the antiferromagnetic (AFM) spin with antiferromagnetic orbital (AFO) (AFM$\times$AFO) phase per bond (in units of $JS^2$) is 
\begin{equation}
   \epsilon_{\rm AFM \times AFO} =- S^2 \left(\frac{3a-2}{3} +\frac{b}{S^2} \right) \times \left( 1+ \frac{b}{S^2}\right).
\end{equation}
Comparing with the energy of the SS$\times$AFO phase:
\begin{equation} \label{eq_E_SS}
   \epsilon_{\rm SS \times AFO} = - S^2 \left(\frac{a+2}{3} +\frac{b}{S^2} \right) \times \left( 1+ \frac{b}{S^2}\right),
\end{equation}
we find that the transition of the two ordered states at $a=2$ is due to the competition between the Kugel-Khomskii term $a(\boldsymbol{S}_i\cdot\boldsymbol{S}_j)(\boldsymbol{T}_i\cdot\boldsymbol{T}_j)$ and the Yao-Lee term $2(S^\gamma_i\cdot S^\gamma_j)(\boldsymbol{T}_i\cdot\boldsymbol{T}_j)$, which is independent of $b$. In the ${\rm SS \times AFO}$ phase, the energy from the Yao-Lee term is maximized, and the Kugel-Khomskii term is partially satisfied (energy $-a$ for two bonds, while $+a$ for the remaining). In contrast, in the ${\rm AFM \times AFO}$ phase, the Kugel-Khomskii term is maximally satisfied, while the Yao-Lee term is unsatisfied. 
To analyze the phase transition in the quantum phase diagram, one must evaluate the spin-wave corrections to the energies of the two competing phases, which are beyond the scope of the current study.

\section{Classical Monte Carlo simulation with parallel tempering} \label{apdx_B}

Classical Monte Carlo simulation (CMC) with simulated annealing is a versatile method for finding the ground-state energy of model Hamiltonians. In simulated annealing, we start at a high temperature and gradually lower it. At each temperature, the system is brought to thermal equilibrium with Metropolis updates \cite{metropolis1953equation, hastings1970monte}. At high temperatures, the system can equilibrate quickly with a few updates, but it undergoes a critical slowdown near a phase transition, thereby increasing the likelihood of reaching metastable states. This problem worsens for models with many competing states near the ground state, multiple phase transitions, or a large lattice size. 

Parallel tempering \cite{Hukushima1996} can accelerate the equilibration by simulating multiple replicas of the system simultaneously, each at a different temperature, and allowing replicas at different temperatures to swap. Parallel tempering allows replicas to be simulated at higher temperatures, where the system can more easily escape metastable states and subsequently relax at lower temperatures on much shorter time scales than simulated annealing. The swap probability that satisfies the detailed balance condition is given by
\begin{equation}
    p(E_i,T_i \rightarrow E_{i+1},T_{i+1}) = \mathrm{min}(1, e^{\Delta\beta\Delta E}),
\end{equation}
where $\Delta\beta=1/T_{i+1}-1/T_i$ is the difference in the inverse temperatures and $\Delta E=E_{i+1}-E_i$ is the difference in the energies.

The key parameters affecting equilibration efficiency are the simulation temperatures of the replicas. For simple models with small lattice sizes, it suffices to use a geometric series of temperature steps that are sufficiently close to each other to allow a replica-exchange rate of approximately 20\% \cite{kone2005selection}. 
For complex models with large lattice sizes, it is desirable to evaluate equilibration quality and further speed up the process using a general feedback-optimized algorithm that selects the optimal temperature steps \cite{Katzgraber2006}. 
We follow Ref. \cite{HAMZE2010} to further improve the convergence stability of this feedback-optimized algorithm by applying a smoothing factor. The algorithm adapts the set of temperatures that ensures replicas can efficiently go round-trips between the highest and lowest temperatures. Physically, the algorithm typically increases the density of temperature steps near a phase transition, thereby alleviating the critical slowdown. The wider the temperature window and the larger the system size, the more iterations are required to find the optimal temperature steps.

For the CMC simulations in this work, we initialize the feedback-optimized algorithm with 200 temperature steps, exponentially increasing from $T=0.001 \mathrm{\tilde{K}}$ to $T=0.1\mathrm{\tilde{K}}$, where $\mathrm{\tilde{K}}=k_BT/J$. We do not use a lower temperature due to time constraints: it would take too long to find the optimal temperature steps that ensure sufficient sampling. 50000 replica exchanges are performed with 5 Metropolis sweeps ($N$, number of unit cells, Metropolis updates each sweep) between exchanges. About 100 iterations of temperature optimization are performed. At each iteration, the number of temperature steps increases, and the number of replica exchanges also increases proportionally. 

\section{generalized spin wave theory } \label{apdx_C}
Here, we show details of the magnon interactions in the generalized (SWT). The spin, orbital, and spin-orbital operators are expressed as linear combinations of the $SU(4)$ generators \cite{Li_SU4_1998PRL, Joshi_SU4_1999PRB} and then mapped to bosons using the generalized Holstein-Primakoff transformation \cite{onufrieva1985, Joshi_SU4_1999PRB}.
When expanded to the order $\Lambda^{-1}$, the transformation reads
\begin{align} 
    O^1_1 &= \Lambda - \sum_{n=2}^{4}b_n^\dagger b_n, \nonumber\\
    O^1_n &= \sqrt{\Lambda}b_n^\dagger \left( 1- \frac{1}{2\Lambda}\sum_{l=2}^{4}b_l^\dagger b_l + \mathcal{O}(\Lambda^{-2}) \right), \nonumber\\
    O^l_n &= b_n^\dagger b_l, \;\;\;\;\; l,n=2,3,4,
\end{align}
which generates both cubic and quartic terms. Note that the cubic terms are also present at order $\Lambda^{0}$ expansion due to spin interactions like { $O^1_2O^3_2$}, which are ignored in the linear SWT Hamiltonian.

To consider the cubic and quartic terms, one needs to go to order $\Lambda^{-1}$.
The quartic terms are treated with the mean-field Hartree-Fock method. For example, 
\begin{align}
    a^\dagger_{k} b^\dagger_{k^\prime} a_q c_{q^\prime} \approx &
    \langle a^\dagger_{k} b^\dagger_{k^\prime}\rangle a_q c_{q^\prime}+
    a^\dagger_{k} b^\dagger_{k^\prime} \langle a_q c_{q^\prime} \rangle+
    \langle a^\dagger_{k} a_q \rangle b^\dagger_{k^\prime}  c_{q^\prime}
    \nonumber \\
    &+a^\dagger_{k} a_q \langle b^\dagger_{k^\prime} c_{q^\prime} \rangle+
    \langle a^\dagger_{k} c_{q^\prime} \rangle b^\dagger_{k^\prime} a_q+
    a^\dagger_{k} c_{q^\prime} \langle b^\dagger_{k^\prime} a_q \rangle 
    \nonumber \\
    \approx & \langle a^\dagger_{q} b^\dagger_{-q}\rangle a_k c_{k}+
    \langle a_q c_{-q} \rangle a^\dagger_{k} b^\dagger_{k} +
    \langle a^\dagger_{q} a_q \rangle b^\dagger_{k}  c_{k}
    \nonumber \\
    &+\langle b^\dagger_{q} c_{q} \rangle a^\dagger_{k} a_k +
    \langle a^\dagger_{q} c_{q} \rangle b^\dagger_{k} a_k+
    \langle b^\dagger_{q} a_q \rangle a^\dagger_{k} c_{k},
\end{align}
where $a^\dagger$,$b^\dagger$,and $c^\dagger$ are three different bosons, and $k+k^\prime-q-q^\prime=0$, $\langle a_k^\dagger a_{k^\prime} \rangle \propto \delta_{k,k^\prime}$, $\langle a_k^\dagger a_{k^\prime}^\dagger \rangle \propto \delta_{k,-k^\prime}$ have been used. 
The quartic terms modify the non-interacting Hamiltonian, leading to the renormalized spectra $\tilde{E}(k)$.

\begin{figure}
    \centering
    \includegraphics[width=0.95\linewidth]{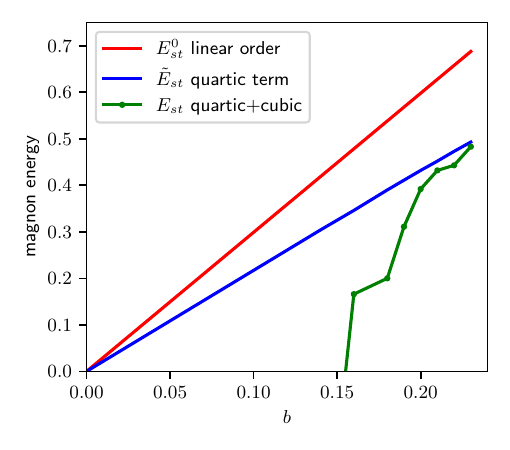}
    \caption{The bare energy $E^0_{st}$, the renormalized energy $\tilde{E}_{st}$, and the corrected energy $\tilde{E}-\Delta E$ of the spin-orbital excitations versus $b$ at $\boldsymbol{k=M}$ and $a=1.0$. The quartic terms simply renormalize the energy and do not change the gap closing at $b=0$. On the other hand, the cubic corrections lead to the gap closing at a much larger $b$, suggesting the importance of the cubic terms for the extended nematic phase.}
    \label{fig_S1}
\end{figure}

The cubic terms are treated with second order perturbation theory to the non-interacting magnon states \cite{Rastelli_1985, Chubukov_1994, Chernyshev2009PRB}. We first diagonalize the quadratic Hamiltonian and transform the $b_n$ operators to quasiparticles $\alpha_n(k)$. After the transformation, we get two types of terms:
\begin{equation}
    \mathcal{H}_3 = C_{k,q}(\alpha,\beta,\gamma)\alpha_k \beta_q^\dagger \gamma_{k-q}^\dagger + D_{k,q}(\alpha,\beta,\gamma) \alpha_k^\dagger \beta_q^\dagger \gamma_{k+q}^\dagger + \mathrm{h.c.},
\end{equation}
where $\alpha,\beta,\gamma$ are three different quasiparticles. 
At order $\Lambda^{-1}$, they generate the normal self-energy and contribute to the correction of the magnon energy given by
\begin{align}
    \Delta E_{\alpha}(k) =& \sum_q \frac{|C(\alpha,\beta,\gamma)|^2}{E_\alpha(k)-E_\beta(q)-E_\gamma(k-q)} \nonumber \\
    &- \sum_q \frac{|D(\alpha,\beta,\gamma)|^2}{E_\alpha(k)+E_\beta(q)+E_\gamma(k+q)}.
\end{align}
Combining the quartic terms, the corrected spectra are $E(k)=\tilde{E}(k)-\Delta E(k)$.
The numerical integrals are estimated using a uniform grid of $\sim4000$ points in the Brillouin zone. 
Figure \ref{fig_S1} shows the bare energy $E^0_{st}$, the renormalized energy $\tilde{E}_{st}$, and the corrected energy $\tilde{E}(k)-\Delta E(k)$ of the spin-orbital excitations versus $b$ at $\boldsymbol{k=M}$, $a=1.0$. The phase transition is determined by the critical value $\tilde{b}_c$ where $E=0$. The quartic terms merely renormalize the interactions and leave the gap‑closing point at $b=0$ unchanged, whereas the cubic terms shift the gap‑closing to a much larger value of $b$.

\bibliography{References}

\end{document}